\documentclass[superscriptaddress,twocolumn,nofootinbib]{revtex4}
\usepackage[T1]{fontenc}
\usepackage{amstext,amsmath,amssymb,amsthm}
\usepackage{color}

\usepackage{float} 

\usepackage{times}

\usepackage{graphicx}

\theoremstyle{plain}

\theoremstyle{definition}

\newcommand{\beq}{\begin{equation}}
\newcommand{\eeq}{\end{equation}}
\newcommand{\ket} [1] {\vert #1 \rangle}
\newcommand{\bra} [1] {\langle #1 \vert}

\newcommand{\ba}{\begin{align}}
\newcommand{\ea}{\end{align}}
\newcommand{\bea}{\begin{eqnarray}}
\newcommand{\eea}{\end{eqnarray}}

\renewcommand{\P}{\mathbb{P}}

\renewcommand{\>}{\rangle}

\makeatletter

\@ifundefined{textcolor}{}
{
 \definecolor{BLACK}{gray}{0}
 \definecolor{WHITE}{gray}{1}
 \definecolor{RED}{rgb}{1,0,0}
 \definecolor{GREEN}{rgb}{0,.6,0}
 \definecolor{BLUE}{rgb}{0,0,1}
 \definecolor{CYAN}{cmyk}{1,0,0,0}
 \definecolor{MAGENTA}{cmyk}{0,1,0,0}
 \definecolor{YELLOW}{cmyk}{0,0,1,0}
 }

\makeatother

\usepackage{bm}
\usepackage{mathtools}

\def\id{I}

\def\1{\mat{\id}}

\def\mat#1{\mathbf{#1}}

\def \BQP {\bf{BQP}}
\def \BPP {{\bf{BPP}}}
\def \MA {\bf{MA}}
\def \NP {\bf{NP}}

\def \P {\bf{P}}

\def \FH {{\bf{FH}}}

\def \inp {\ket{{\bf s}_{\text{in}}}}
\def \sinp {{\bf s}_{\text{in}}}

\begin{document}
\title{Classical verification of quantum circuits containing few basis changes}
\author{Tommaso F. Demarie}\email{tommaso_demarie@sutd.edu.sg}
\affiliation{Singapore University of Technology and Design, 8 Somapah Road, Singapore 487372}

\author{Yingkai Ouyang}\email{yingkai_ouyang@sutd.edu.sg}
\affiliation{Singapore University of Technology and Design, 8 Somapah Road, Singapore 487372}

\author{Joseph F. Fitzsimons}\email{joseph_fitzsimons@sutd.edu.sg}
\affiliation{Singapore University of Technology and Design, 8 Somapah Road, Singapore 487372}
\affiliation{Centre for Quantum Technologies, National University of Singapore, 3 Science Drive 2, Singapore 117543}

\begin{abstract}
We consider the task of verifying the correctness of quantum computation for a restricted class of circuits which contain at most two basis changes. This contains circuits giving rise to the second level of the Fourier Hierarchy, the lowest level for which there is an established quantum advantage. We show that, when the circuit has an outcome with probability at least the inverse of some polynomial in the circuit size, the outcome can be checked in polynomial time with bounded error by a completely classical verifier. This verification procedure is based on random sampling of computational paths and is only possible given knowledge of the likely outcome.
\end{abstract}

\maketitle

\emph{Introduction} --- One of the great puzzles of quantum computation is the origin of its apparent power over classical computation. Among different attempts to find a solution to this quandary, it has been suggested that the answer might lie in the particular structure of the computation. In particular, it was proposed in~\cite{Shi2005} that the \emph{quantumness} of a quantum circuit is derived from layers of operations that do not preserve the computational basis. The intuition behind this claim follows from a simple but insightful observation. Any $n$-qubit unitary operation can be approximated to an arbitrary level of accuracy by using gates from certain finite \emph{approximately universal} gate sets. One particular approximately universal set contains only two types of gate: The Toffoli gate and the Hadamard gate~\cite{Shi2003,Aharonov2003}. The Toffoli gate is universal for (reversible) classical computation, and in quantum circuits it has an entirely classical flavour since it preserves the computational basis. This seems to indicate that the quantum advantage is introduced by the gates that \emph{do not} preserve the computational basis, i.e. the Hadamard gates. With this view in mind, the Fourier hierarchy ${\bf FH}$ was introduced in~\cite{Shi2005}. Each level ${\bf FH}_k$ of the hierarchy corresponds to the class of problems solvable by polynomial-size quantum circuits, composed of gates that preserve the computational basis and $k$ layers of operations that do not preserve it.

In this work we prove that circuits containing up to two Fourier transforms which produce likely outcomes can be efficiently verified by an entirely \emph{classical} computer. With this in mind, we will use $\FH_2$ to denote both decision problems and the class of circuits containing at most two Fourier transforms, with the meaning clear from the context. Importantly, $\FH_2$ contains circuits that exhibit clear advantages over their classical counterparts. The paradigm of \emph{verification of quantum computation} lies deep into the roots of quantum mechanics, raising questions about the falsifiability of the theory in regimes of high computational complexity \cite{Aharonov2013}. The challenge is to certify the result of a quantum computation using devices that are themselves unable to derive that result. This is an issue that is not only of theoretical interest. Developments in the experimental control of quantum systems in the last decade has increased the difficulty of verifying the consistency of an experiment's outcome with regards to the predictions of quantum mechanics. While the simulation of the quantum evolution of systems comprising of a small number of qubits on a classical computer is possible, the difficulty of this simulation grows exponentially with the size of the quantum computer. Hence one requires new techniques to solve the problem of verification. Recent claims about the \emph{quantumness} of a certain types of experimental processors~\cite{Boixo2013,Boixo2014} have sparked both excited reactions and strong criticisms~\cite{Smolin2013, Shin2014b, Shin2014} and more generally caused a passionate debate~\cite{Wang2013, Zintchenko2015, Denchev2015} that suggests how coming up with a feasible approach for the verification of quantum computation is of practical importance.

These issues have motivated recent theoretical efforts to develop novel protocols for quantum verification. Generally, these protocols are presented as interactive games where a \emph{verifier} with limited computational resources attempts to verify the output of a quantum computation performed by a \emph{prover} capable of processing quantum information. Such verification protocols rely on different methods: The embedding of various types of veracity tests~\cite{Fitzsimons2012, Barz2013,Morimae2014ver,Hayashi2015,Broadbent2015} into blind quantum computing protocols \cite{Broadbent2009,Barz2012,Morimae2013b}, approaches based on self-testing~\cite{McKague2013,Reichardt2013,Reichardt2013bis}, hybrid techniques combining these two procedures~\cite{Gheorghiu2015, Hajdusek2015} and variety of methods based on the use of error correction codes~\cite{Aharonov2010,Fitzsimons2015,Morimae2016}. A common thread, however, is the need for at least two parties with quantum capabilities: either a verifier with limited quantum capabilities or multiple quantum provers sharing entanglement. While a program to explore classically driven blind quantum computing was initiated in~\cite{Mantri2016}, it remains an open question whether decision problems in $\BQP$ can be efficiently verified by a prover \emph{without any} quantum power. 

Here, we explore the possibility of verifying a single quantum processor using purely classical means, restricting ourselves to quantum circuits that belong to $\FH_2$, the second level of the Fourier Hierarchy. In particular, we focus on quantum computations that have likely outcomes. This is motivated by considerations of \emph{usefulness}: Quantum algorithms that are believed to offer an advantage over their classical analogs, such as factoring~\cite{Shor1994} and quantum search~\cite{Grover1996} algorithms, are designed to deliver the correct answer with high probability. On the other hand, models of quantum computation based on sampling are not known to have practical applications \footnote{see for example the the discussion on boson sampling in~\cite{Gard2015}}. We therefore exploit the structure of $\FH_2$ to show that a polynomial-time classical verifier can efficiently verify the outcome of a $\FH_2$ circuit implemented by a prover, with only a single round of communication between them. In analogy with Ref.~\cite{Fitzsimons2015}, this proof does not rely on blindness, and is suggestive of the possibility that $\FH_2 \subseteq \MA$, a possibility made more interesting since it is not yet known whether $\BQP = \FH_2$.

We begin with some terminology. If ${\bf s}=(s_1,\dots, s_n)$ is an $n$-bit string, we denote by $\ket{{\bf s}} = |s_1\> \otimes \dots \otimes |s_n\>$ the corresponding computational basis state. A reversible classical computation $C$ is a bijection from $n$-bit strings to $n$-bit strings. We consider the corresponding quantum circuits $\hat C$ that are bijections from $n$-qubit computational basis states to $n$-qubit computational basis states, and say that such quantum circuits are \emph{classical}. Since each such classical transformation is a permutation on the $2^n$ computational basis states, the set of all such circuits $\mathcal{P}_C$ is isomorphic to the symmetric group on $2^n$ symbols \cite{Shepherd2010}. We call $\mathcal{P}_C$ the permutation group on the computational basis, and it can be generated by the set of generalised $k$-Toffoli gates, where $k$ indicates the number of control qubits (i.e. for $k=0$ we have a Pauli-$X$, for $k=1$ a CNOT gate and so on).

Gates that do not preserve the computational basis naturally extend the permutation group on the computational basis. When a gate $\hat G$ has such a property, there necessarily exists some computational basis elements $\ket{{\bf i}}$ and $\ket{{\bf j}}$ such that $0 < |\bra{{\bf i}} \hat{G} \ket{{\bf j}} | <1$. We call such gates \emph{basis-changing} gates. The simplest example of a basis-changing gate for a single qubit is the Hadamard gate $\hat{H}$, which plays the role of a quantum Fourier transform~\cite{Nielsen2000} by rotating the vectors $\ket{0}$ and $\ket{1}$ onto the perpendicular $(X,Y)$-plane of the Bloch sphere. In general, the quantum Fourier transform on $n$ qubits can be implemented by a poly-$(n)$ combination of Hadamard gates and $\frac{\pi}{8}$-gates~\cite{Nielsen2000}. Since any quantum circuit can be approximated by a sequence of Toffoli and Hadamard gates, one can think of quantum circuits as procedures that alternate between classical (Toffoli) and quantum (Hadamard) information processing. This line of thought leads directly to the \emph{Fourier Hierarchy}. Given a non-negative integer $k$, $\FH_k$ is the complexity class of problems that can be decided with bounded error probability by quantum circuits of polynomial size containing classical gates and at most $k$ quantum Fourier transforms. The Fourier hierarchy captures part of the subtlety of quantum computation, and its lowest levels correspond to some of the most common complexity classes, which are informally introduced hereafter. Rigorous definitions can be found in~\cite{Watrous2008}.

A decision problem deterministically answerable by a classical computer within time polynomial in the input size belongs to the complexity class {\bf P}. The class {\bf NP} corresponds to decision problems for which \emph{yes} instances can be deterministically verified in polynomial time by a classical computer, given a suitable witness string, and so trivially {\bf P} $\subseteq$ {\bf NP}. If a classical computer, augmented with the ability to generate randomness, can instead answer a decision problem with error probability bounded by some constant $\lambda <\frac{1}{2}$ in polynomial time, this decision problem is contained in the class {\bf BPP}. Both $\NP$ and $\BPP$ are contained in a class known as $\MA$. A decision problem belongs to $\MA$ if it has a witness string which can be verified by a polynomial time verifier with bounded probability of error. The class $\MA$ differs from {\bf NP} because in $\MA$ the verifier has a bounded non-zero probability to accept a {\em no instance}.

Moving from classical to quantum devices, {\bf BQP} is the complexity class corresponding to decision problems that can be answered with bounded error probability by a \emph{quantum} computer in polynomial time. If a {\em yes} instance of a decision problem can be verified with bounded probability of error by a quantum polynomial time verifier with the aid of a particular quantum \emph{proof} state, that decision problem belongs to the class {\bf QMA}. The hierarchical relations {\bf P} $\subseteq$ {\bf BPP} $\subseteq$ {\bf BQP} $\subseteq$ {\bf QMA}, and {\bf NP} $\subseteq \MA \subseteq {\bf QMA}$ hold. Note that the relationship between {\bf NP} and {\bf BQP} is unknown, although it is conjectured that {\bf NP} $\nsubseteq$ {\bf BQP} and that {\bf BQP} $\nsubseteq$ {\bf NP} \cite{aaronson2010bqp}.
 
Let us allow only uniform families of quantum circuits. Then, given the definitions above, it is easy to see that $\FH_0 = {\bf P}$: Any decision problem represented as a quantum circuit composed solely of classical gates corresponds to a decision problem in $\P$. It also follows that $\FH_1 = \BPP$ since, for an input state in the computational basis, a single change of basis cannot cause phase interference. This means that, for a computational basis input, the quantum output of a $\FH_1$ circuit is uniformly distributed on the support of the Fourier transform, and it gives access to randomness elevating $\P$ to $\BPP$. Characterising the levels of the Fourier hierarchy becomes intriguing in terms of complexity for $k \ge 2$. Kitaev's phase estimation algorithm~\cite{Kitaev1995} can be used to derive an efficient quantum algorithm for integer factoring that requires two Fourier transforms.
Therefore, Shor's algorithm~\cite{Shor1994} for factorisation, which gives a substantial speedup when compared to the most efficient known classical algorithm for factorisation, belongs to $\FH_2$. One might then wonder if two layers of quantum Fourier transforms, or basis-changing gates in general, suffice to unlock the power of quantum computation. To date, an exact relationship between $\FH_2$ and the other complexity classes remains unknown.

Our main result deals with the verification of circuits in $\FH_2$, that is quantum circuits with two layers of Fourier transforms preceded, interspaced, and followed by classical computation from $\mathcal{P}_C$ composed of a number of gates polynomial in the input size. Consider a prover performing the circuit just described on a generic input in the computational basis. 
The prover claims that the classical outcome of the computation, after measuring the quantum state obtained at the end of the circuit in the computational basis, is the $n$-bit string ${\bf s}=(s_1,\dots, s_n)$. 
The verification problem we consider is to decide whether the probability of obtaining {\bf s} is large or alternatively small, \emph{under the promise} that \emph{exactly one} of these two instances holds and that their separation is at most some inverse polynomial in $n$. We prove that the verification process can be performed by a randomized polynomial time classical verifier with access to the classical description of the input state, the quantum circuit and the string ${\bf s}$.

We begin by giving a definition of the class of basis-changing gates used in the quantum circuits that we consider. We will say that an $n$-qubit unitary operator $\hat T$ is a {\em classical samplable transform} if it satisfies the following set of conditions: 
\begin{enumerate}
\item $\hat{T}$ can be implemented by a number of Toffoli, Hadamard and $\frac{\pi}{8}$-gates polynomial in the input size $n$.
\item For all ${{\bf s}_1} \in \{0,1\}^n$, there exists a polynomial time randomised classical algorithm which randomly samples a distribution over $n$ bit strings such that the probability of outputting ${{\bf s}_2} \in \{0,1\}^n$ is
\begin{equation}
p_{{\bf s}_2}^{{\bf s}_1} = \frac{|\bra{{{\bf s}_2}}\hat{T}\ket{{{\bf s}_1}}|}
{\sum_{{\bf s} \in \{0,1\}^{n}} |\bra{{\bf s}}\hat{T}\ket{{{\bf s}_1}}|}.
\end{equation}
\item For every ${{\bf s}_1}$ and ${{\bf s}_2}$, the complex phase of $\bra{{{\bf s}_2}}\hat{T}\ket{{{\bf s}_1}}$, can be computed in classical polynomial time.
\end{enumerate}
Any tensor product of the identity operator, Hadamard transforms, and Fourier or inverse Fourier transforms on disjoint systems satisfies the above definition. 
Let $S_{\hat{F}} \subseteq \{1, ...,n\}$. Then, we say that $S_{\hat{F}}$ is the support of $\hat{F}$ if $\hat{F}$ acts non-trivially on the qubits labelled by the elements of $S_{\hat{F}}$. Given an input state $\ket{{\bf s}}=(s_1, ..., s_n)$ we use $\mathcal{B}(\hat{F},\ket{\bf s})$ to denote the set of all $n$-bit strings where the $i$-th component is equal to $s_i$ for all $i \notin S_{\hat{F}}$. It follows that such operations have the property that $p_{{{\bf s}_2}}^{{{\bf s}_1}} = \frac{1}{2^{m}}$, where $m$ is the cardinality of $S_{\hat{F}}$. We shall restrict our attention to classically samplable transforms for which this is true. We thereby define a \textit{$k$-transform circuit}, which is a quantum circuit $\mathcal{C}$ that has the following properties.
\begin{enumerate}
\item The input to $\mathcal{C}$ is a computational basis state.
\item The quantum circuit $\mathcal{C}$ comprises of a polynomial number of Toffoli gates (basis preserving) and $k$ classically samplable transforms (basis changing), followed by measurement of all qubits in the computational basis.
\item The output of $\mathcal{C}$ is the bit string that corresponds to the measured computational basis state.
\end{enumerate}

Having defined the circuits under examination, we cast the corresponding verification task as a decision problem with the promise that the input satisfies the requirements for either a {\em yes} instance or a {\em no} instance as we now describe. We say that a $k$-transform circuit is \textit{$\delta$-deterministic} with output ${\bf s}$ if the measurement outcome after running the circuit is ${\bf s}$ with probability at least $\delta$. In the $k$-transform verification problem, an instance consists of a $k$-transform circuit $\mathcal{C}$ and a string $\bf s$, with the promise that exactly one of the following instances is true.
\begin{enumerate}
\item The {\em yes} instance: $\mathcal{C}$ is $\delta$-deterministic with output $\bf s$.
\item The {\em no} instance: $\mathcal{C}$ is not $\epsilon$-deterministic for any output. 
\end{enumerate} 
The task is to decide if either the {\em yes} instance or the {\em no} instance holds for the circuit $\mathcal C$, where $\delta$ and $\epsilon$ are defined as follows. Both $\delta$ and $\epsilon$ are positive real numbers in the interval $[0,1]$ such that $\epsilon < \delta/2$, and $\gamma = \sqrt{\frac{\delta}{2}} - \sqrt{\epsilon}$ satisfies $\gamma = \Omega(\text{poly}^{-1}(n))$. This last constraint is required ensure that the probabilities are sufficiently distinct so that the difference can be resolved with a polynomial number of samples.

Our main result is that the $k$-transform verification promise problem is in {\BPP} for $k\leq 2$. It suffices to show that if $\mathcal{C}$ is $\delta$-deterministic then there exists a proof of this fact that can be verified by a classical prover in polynomial time with bounded error of $\frac{1}{3}$, and that this verification procedure rejects any proof with bounded error of $\frac{1}{3}$ if $\mathcal{C}$ is not $\epsilon$-deterministic. For the proof we use the structure of $\FH_{k}$ for $k\le 2$. In particular, $\FH_0 = \P$, and $\FH_1 = \BPP$. When $k=0$, the circuit is completely classical, and hence it can be verified by direct evaluation. When $k=1$, consider the following argument. Let us call each layer of classical computation $\hat{C}_i$, where the index $i$ indicates the temporal order of the layer in the circuit. Then the output state of $\mathcal{C}$ before the final measurement is $\hat{C}_2 \hat{F}_1 \hat{C}_1 \ket{\sinp}$ with an $n$-qubit computational basis input state $\ket{\sinp}$. Here $\hat{C}_1$ and $\hat{C}_2$ are polynomial sized Toffoli circuits in $\mathcal P_C$, and $\hat{F}_1$ is a classically samplable transform. Note that $C_1 ({\bf s}_{\text{in}}) = {\bf r}$ for some $n$-bit sting {\bf r} and hence $\hat{C}_1 \inp = \ket{C_1 (\sinp)} = \ket{{\bf r}}$. Because of the reversible classical property of $\hat{C}_2$, the verifier can efficiently derive $|C_2^{-1}(\bf s)\rangle$, where $\hat{C}_2 \ket{C_2^{-1}({\bf s})} = \ket{{\bf s}}$. Finally the complex phase $\bra{C_2^{-1}({\bf s})} \hat{F}_1 \ket{\bf r}$ can be trivially computed by definition. This answers the verification problem for $k=1$.

We now evaluate the probability that a fixed output string {\bf s} is obtained from any $2$-transform circuit evaluated on the $n$-qubit computational basis state $\ket{{\bf s}_{\rm in}}$. The output of a $2$-transform circuit $\mathcal{C}$  before the measurement can be written as $\hat{C}_3 \hat{F}_2 \hat{C}_2 \hat{F}_1 \hat{C}_1 \ket{{\bf s}_{\text{in}}}$ where the transforms $\hat{F}_1,\hat{F}_2$ act non-trivially on $a\le n$ and $b \le n$ qubits respectively. Then
\begin{equation}
\hat{F}_1 \ket{{\bf r}} = 2^{-\frac{a}{2}} \sum_{{\bf j} \in \mathcal{B}(\hat{F}_1, \ket{\bf r})}
 e^{i \alpha_{{\bf r},{\bf j}}} \ket{{\bf j}}\,,
\end{equation}
where $\alpha_{{\bf r},{\bf j}}$ is the phase for the complex amplitude of the state $\ket{{\bf j}}$ produced by the samplable transform given the fixed input $\ket{{\bf r}}$. Then
\begin{equation}
\hat{C}_2 \hat{F}_1 \ket{{\bf r}} = 2^{-\frac{a}{2}} \sum_{{\bf j} \in \mathcal{B}(\hat{F}_1, \ket{\bf r})}e^{i \alpha_{{\bf r},{\bf j}}} \ket{C_2({\bf j})}\, ,
\end{equation}
and
\begin{equation}
\hat{F}_2 \hat{C}_2 \hat{F}_1 \ket{{\bf r}} 
= 
2^{-\frac{a+b}{2}}
  \sum_{
  \substack{
  	{\bf j} \in \mathcal{B}(\hat{F}_1, \ket{\bf r})\\
  	{\bf k} \in \mathcal{B}(\hat{F}_2, \ket{C_2(\bf j)})
   }}
  e^{i \alpha_{{\bf r},{\bf j}}} e^{i \beta_{C_2({\bf j}),{\bf k}}} \ket{{\bf k}} \, ,
\end{equation}
where each $\beta_{C_2({\bf j}),{\bf k}}$ is the phase associated to the complex amplitude of each state $\ket{\bf k}$ induced by the action of $\hat{F}_2$ on the state $\ket{C_2(\bf j)}$. The combined action $\hat{F}_2 \hat{C}_2 \hat{F}_1$ makes the computation difficult to simulate classically using known techniques. This form, equivalent to the core of Shor's algorithm, likely cannot be simulated efficiently by a classical circuit because the gate $\hat{C}_2$ is performed on a superposition of computational basis vectors~\cite{Aharonov2007}. Indeed, such circuits allow for the preparation and measurement in the $XY$-plane and $Z$-basis of arbitrary graph states, and hence can be used to implement uncorrected measurement-based computation \cite{Raussendorf2003}. Under post-selection this becomes universal, and hence by standard arguments \cite{bremner2010classical,morimae2014hardness,rohde2015evidence} sampling the output of 2-transform circuits within bounded multiplicative error is computationally hard classically. However, with knowledge of $\bf s$, Born's rule $P_{{\bf s}} = | \bra{C_3^{-1}({\bf s})} \hat{F}_2 \hat{C}_2 \hat{F}_1 \ket{{\bf r}} |^2$ gives the probability of obtaining the output {\bf s}, which can be estimated using a sampling technique as follows.

A randomised classical sampling algorithm that runs in a time polynomial in $n$ is used to answer the verification problem for any $2$-transform circuit on $n$ qubits. To show this, we start with the amplitude $\xi_{\bf s}=\bra{C_3^{-1}({\bf s})} \hat{F}_2 \hat{C}_2 \hat{F}_1 \ket{{\bf r}}$ associated to the state $\ket{\bf s}$. One needs to distinguish between the $b\ge a$ and $a > b$ cases. In the following we will only consider the former case, since the same analysis can be performed for the latter case by first taking the complex conjugate of  the amplitude $\xi_{\bf s}$ and expanding over paths through $\hat{F}_2$ rather than $\hat{F}_1$, as is done next. We expand the amplitude as
\begin{align*}
\xi_{\bf s} &= \nonumber
	2^{-\frac{a}{2}} 
		\sum_{{\bf j} \in \mathcal{B}(\hat{F}_1, \ket{\bf r})} 
			 e^{i \alpha_{{\bf r},{\bf j}}} 
				\bra{C_3^{-1}({\bf s})} \hat{F}_2 \ket{C_2({\bf j})} \\ \nonumber
	&=2^{-\frac{a+b}{2}} 
		\sum_{{\bf j} \in \mathcal{B}(\hat{F}_1, \ket{\bf r})}
			 \theta_{C_2({\bf j}), C_3^{-1}({\bf s})}
				e^{i \alpha_{{\bf r},{\bf j}} + i \beta_{C_2({\bf j}),{C_3^{-1}({\bf s}) }}},
\end{align*}
where $\theta_{C_2({\bf j}), C_3^{-1}({\bf s})} \in \{0,1\}$ depending on whether $\bra{C_3^{-1}({\bf s})} \hat{F}_2 \ket{C_2({\bf j})}$ is non-zero. To simplify notation, we define
\begin{align*}
	u_{\bf j} &= 2^{-a}\text{Re}\left(\theta_{C_2({\bf j}), C_3^{-1}({\bf s})} e^{i \alpha_{{\bf r},{\bf j}} + i \beta_{C_2({\bf j}),{C_3^{-1}({\bf s}) }}}\right) \, , \text{ and}\\
	v_{\bf j} &= 2^{-a}\text{Im}\left(\theta_{C_2({\bf j}), C_3^{-1}({\bf s})} e^{i \alpha_{{\bf r},{\bf j}} + i \beta_{C_2({\bf j}),{C_3^{-1}({\bf s}) }}}\right) \, ,
\end{align*} 
so that $\xi_{\bf s} = 2^{-\frac{(b-a)}{2}}\left(\sum_{\bf j} u_{\bf j} + i v_{\bf j} \right)$. Note that $2^{-\frac{b-a}{2}} \ge |\xi_{\bf s}|$, which implies that all the cases where $b-a = \Omega(\text{poly}(n))$ are trivial to analyse, since they cannot be $\text{poly}^{-1}(n)$-deterministic for any $\bf s$. In the following we use the rescaled values $\delta'= 2^{b-a} \delta$ and $\epsilon' = 2^{b-a} \epsilon$ such that $\gamma' =  \sqrt{\frac{\delta'}{2}} - \sqrt{\epsilon'}$. Let $A=2^{-a} \sum_{{\bf j} \in \mathcal{B}(\hat{F}_1, \ket{\bf r})} u_{\bf j}$ and $B=2^{-a} \sum_{{\bf j} \in \mathcal{B}(\hat{F}_1, \ket{\bf r})} v_{\bf j}$. It follows that when $|\xi_{\bf s}|^2 \ge \delta$ we have $|A + i B|  \ge  \sqrt{\delta'}$, then either $|A| \ge \sqrt{\frac{\delta'}{2}}$ or $|B| \ge \sqrt{\frac{\delta'}{2}}$ is true. When $|\xi_{\bf s}|^2 \le \epsilon$, from the triangle inequality, the inequality $|A + i B| \le \sqrt{\epsilon'}$ implies that both $|A| \le \sqrt{\epsilon'}$ and $|B| \le \sqrt{\epsilon'}$ are true.

Using the variables $u_{\bf j}$ and $v_{\bf j}$ we define the independently and identically distributed random variables $\hat X_i$ for $i = 1,\dots, N$ where $N$ is polynomial in $n$ and $\Pr(\hat X= u_{\bf j} + i v_{\bf j} ) = 2^{-a}$ for all ${\bf j} \in \mathcal{B}(\hat{F}_1, \ket{\bf r})$. The definition of the classically samplable transform ensures that there exists a polynomial time randomised classical algorithm for sampling the set $\{ \hat{X}_{i} \}_{i=1}^N$. Let $\hat A$ and $\hat B$ be the real and imaginary parts of $\frac{1}{N} \sum_i \hat X_i$ respectively. Let $\theta = \sqrt{\epsilon'} + \gamma'/2$. Without loss of generality assume that at the end of the sampling $|\hat A| \ge |\hat B|$. If this is the case, when $|\hat A| < \theta$, the verifier concludes that $|A + i B| \le \sqrt{\epsilon'}$, and if $|\hat A| \ge \theta$, the verifier concludes that $|A+iB| \ge \sqrt{\delta'}$ since the promise of the problem excludes the possibility that $\sqrt{\frac{\delta'}{2}} \leq |A+iB| < \sqrt{\delta'}$. If $|\hat A| \le |\hat B|$ the same conclusions apply when substituting $|\hat A|$ with $|\hat B|$. In the following paragraphs we prove that the conclusion of the verifier is incorrect with probability exponentially small in $N$.

Here we make use of the Hoeffding bound \cite{Hoeffding1963} and the reverse triangle inequality applied to probabilities. Hoeffding's bound states that 
$ \Pr\left[  |   \hat A - A | \ge \frac{\gamma'}{2} \right] \le 2e^{-\gamma'^2 N/8}.$
The reverse triangle inequality implies that
$|  \hat A - A| \ge 
| |    \hat A| -  |A||$, and hence
\begin{align}
\Pr \left[
| |  \hat A| -  |A||  \ge \frac{\gamma'}{2}
\right]
\le
\Pr\left[|  \hat A - A|  \ge \frac{\gamma'}{2} \right]
. \label{eq:reverse-triangle-with-probability}
\end{align} Note that when $|A|\ge \sqrt{\delta'/ 2}$,
\begin{align}
\Pr \left[|   \hat A|  \le \theta\right]
\le 
\Pr \left[ |A| -  |   \hat A|  \ge \frac{\gamma'}{2}\right].\label{eq:ahat-theta}
\end{align}
Combining the inequalities in Eq.~\ref{eq:reverse-triangle-with-probability} and Eq.~\ref{eq:ahat-theta} with the Hoeffding bound results in $\Pr [|   \hat A|  \le \theta ] \le 2e^{-\gamma'^2 N/8}.$ When $|A|\le \sqrt{\epsilon'}$,
\begin{align}
\Pr \left[|   \hat A|  \ge \theta\right]
\le 
\Pr \left[ |\hat A| -  |  A|  \ge \frac{\gamma'}{2}\right].
\end{align}
By similar reasoning to the previous case, this yields $\Pr  [|\hat A|  \ge \theta ] \le 2e^{-\gamma'^2 N/8}.$
 
We have hence shown that a randomised classical algorithm can distinguish between the {\em yes} and the {\em no} instance with probability at least $1 -  2 e^{-\gamma'^2 N/8}$. This classical test assesses if the string ${\bf s}$ is a likely outcome of the quantum computation and gives a protocol for the classical verification of a 2-transform circuit $\mathcal{C}$:
\begin{enumerate}
\item The prover performs $\mathcal{C}$. It generates a classical output string ${\bf s}$ and sends it to the verifier. 
\item The verifier uses the string ${\bf s}$ to identify the amplitude $\bra{C_3^{-1}({\bf s})} \hat{F}_2 \hat{C}_2 \hat{F}_1 \ket{{\bf r}}$. It then classically samples $N$ complex phases $\{ \hat{X}_j\}$, with $\hat{X}_j = \hat{A}_j + i \hat{B}_j$.
\item If $|\hat A| > \theta$ and $|\hat B| > \theta$ the verifier accepts the result ${\bf s}$, and it rejects otherwise.
\end{enumerate}
If the circuit $\mathcal{C}$ is $\delta$-deterministic with outcome $\bf s$, the verifier will accept with probability at least $p$ if $N > 8 \gamma^{-2} \log{\frac{2}{1-p}}$, and reject with at least the same probability otherwise.

The fact that the $k$-transform verification problem is in $\BPP$ for $k\le2$ bears relevant consequences. We can modify the question by asking whether there exists any $\bf s'$ for which $\mathcal{C}$ is $\delta$-deterministic, given the promise as before that either such an $\bf s'$ exists, or the circuit is not $\epsilon$-deterministic for any output. Since $\bf s$ acts as a witness for this, using the previous algorithm, it follows that this problem is in {\MA} for $k\le 2$. Furthermore, this witness can be efficiently found by sampling $\mathcal{C}$ with high probability, which can be accomplished by a prover limited to efficient quantum computation.

\section*{Acknowledgements}
The authors acknowledge support from Singapore's National Research Foundation and Ministry of Education. TFD thanks Atul Mantri and Michal Hajdu\v{s}ek for interesting and stimulating discussions. JFF acknowledges support from the Air Force Office of Scientific Research under AOARD grant FA2386-15-1-4082. This material is based on research funded in part by the Singapore National Research Foundation under NRF Award NRF-NRFF2013-01.

\bibliographystyle{bibstyle_notitle_papers}
\bibliography{allrefs-Tom}

\end{document}